\documentclass[12pt,a4paper]{article}
\usepackage[margin=2cm]{geometry}
\usepackage{caption} 
\usepackage{amsmath,amsthm,amssymb,bm,hhline,psfrag,comment}
\usepackage[hyphens]{url}
\usepackage{color}
\usepackage{proof}
\usepackage[dvipdfmx]{graphicx}
\usepackage[authoryear,round]{natbib}

\def\hsymbu#1{\smash{\lower1.7ex\hbox{\huge$#1$}}}

\usepackage{indentfirst}
\usepackage{algorithm,algorithmic}
\usepackage{appendix}
\usepackage[normalem]{ulem}
\usepackage{nomencl}

\makeatletter
\def\eqnarray{\stepcounter {equation}\let \@currentlabel =\theequation
\global \@eqnswtrue
\global \@eqcnt \z@ \tabskip \@centering \let \\=\@eqncr
$$\halign to \displaywidth \bgroup \@eqnsel \hskip \@centering
$\displaystyle \tabskip \z@ {##}$&\global \@eqcnt \@ne \hfil
${\mbox{}##\mbox{}}$\hfil &\global \@eqcnt \tw@
$\displaystyle \tabskip \z@ {##}$\hfil \tabskip \@centering
&\llap {##}\tabskip \z@ \cr}
\makeatother

\theoremstyle{plain}

\theoremstyle{remark}

\theoremstyle{example}

\theoremstyle{lemma}

\theoremstyle{Theorem}

\begin{document}

{
\begin{center}
\textbf{\Large Interpretable modeling for short- and medium-term electricity load forecasting 
}
\end{center}
\begin{center}
\large {Kei Hirose $^{1,2}$ 
}
\end{center}

\begin{flushleft}
{\footnotesize
$^1$ Institute of Mathematics for Industry, Kyushu University, 744 Motooka, Nishi-ku, Fukuoka 819-0395, Japan \\

$^2$ RIKEN Center for Advanced Intelligence Project, 1-4-1 Nihonbashi, Chuo-ku, Tokyo 103-0027, Japan \\
}
{\it {\small E-mail: hirose@imi.kyushu-u.ac.jp}}	
\end{flushleft}

\vspace{1.5mm}

\begin{abstract}
We consider the problem of short- and medium-term electricity load forecasting by using past loads and daily weather forecast information.   Conventionally, many researchers have directly applied regression analysis.  However, interpreting the effect of weather on these loads is difficult with the existing methods.  In this study, we build a statistical model that resolves this interpretation issue.  A varying coefficient model with basis expansion is used to capture the nonlinear structure of the weather effect.  This approach results in an interpretable model when the regression coefficients are nonnegative.  To estimate the nonnegative regression coefficients, we employ nonnegative least squares.  Three real data analyses show the practicality of our proposed statistical modeling. Two of them demonstrate good forecast accuracy and interpretability of our proposed method.  In the third example, we investigate the effect of COVID-19 on electricity loads.  The interpretation would help make strategies for energy-saving interventions and demand response.
 \end{abstract}
 \noindent {\bf Key Words}:  basis expansion; COVID-19; nonnegative least squares; short-term load forecasting; varying coefficient model

 \section{Introduction}
Short- and medium-term load forecasting with high accuracy is essential for decision making during the trade on electricity markets and operation of power systems.  Conventionally, several researchers have used previous loads, weather, and other factors as exploratory variables (e.g., \citealp{Lusis:2017gd}) and directly applied regression analyses to forecast loads.   As methodologies for regression analysis, linear regression \citep{Amral:2008bc,Dudek:2016hy,8285261} and smoothing spline \citep{Engle:1986bb,Harvey:1993} have been traditionally used.   Recently, several studies have applied functional data analysis, where the daily curves of electricity loads are expressed as functions \citep{Cabrera:2017be,Vilar:2018iy}. 
It should be noted that most statistical approaches are based on probabilistic forecasts, and the distribution of forecast values is helpful for risk management \citep{Cabrera:2017be}. On the other hand, machine learning techniques have attracted attention in recent years, such as support vector machine (SVM; \citealp{Chen:2017fy,Jiang:2018bu,yang2019short}), neural networks \citep{He:2017ch,Kong:2018cn,Guo:2018fe,Bedi:2019gc,wang2019probabilistic}, gradient boosting \citep{zhang2019comparison}, and hybrids of multiple forecasting techniques \citep{Miswan:2016ve,Liu:2017fq,deOliveira:2018be,haq2019new}.  These techniques capture complex nonlinear structures; therefore, high forecast accuracies are expected.

 In practice, the time intervals are different among exploratory variables.   For example, assume that loads are forecasted for a single day that will occur several days in the future, at 30-minute intervals; this is a common scenario for market transactions in electricity exchanges (e.g., the day-ahead market in the European Power Exchange, EPEX).  In this example, the electricity loads would be collected at 30-minute intervals using a smart meter, whereas weather forecast information, such as maximum temperature and average humidity, would typically be observed at intervals of one day.  In this study, we use past loads at 30-minute or 1-hour intervals and daily information (e.g., maximum temperature) on the forecast day as exploratory variables.  We note that our proposed model, which will be described in Section 2, is directly applicable to any time resolution of data, such as the load in 1-minute intervals and temperature forecast in 1-hour intervals.

From a suppliers' point of view, it is crucial to produce an interpretable model to investigate the impact of weather on the loads.  For example, estimating the fluctuations of electric power caused by weather would help develop strategies for energy-saving interventions (e.g., \citealp{guo2018residential,wang2018impact}) and demand response (e.g., \citealp{ruiz2020integration}).  To produce an interpretable model, one can directly add weather forecast information to the exploratory variables in the regression model \citep{Hong:2010gl} and investigate the estimator of regression coefficients.   More generally, techniques for interpreting any type of black-box model, including the deep neural networks, have been recently proposed, such as the Local Interpretable Model-agnostic Explanations (LIME; \citealp{ribeiro2016should}) and SHapley Additive exPlanations (SHAP; \citealp{lundberg2017unified}).   However, these methods are used for variable selection, i.e., a set of variables that plays an essential role in the forecast is selected.  Variable selection cannot estimate the fluctuations in electric power caused by weather.

In contrast to variable selection, decomposition of the electricity load at time $t$, say $y_t$, into two parts is useful for interpretation:
\begin{equation}
    y_t \approx \mu_t + b_t, \label{eq:decomposition}
\end{equation}
 where $\mu_t$ and $b_t$ are the effects of past loads and weather forecast information, respectively.   Typically, we use loads at the same time interval of the previous days as exploratory variables (e.g., \citealp{Lusis:2017gd}), and regression analysis is separately conducted on each time interval.   We then construct estimators of $\mu_t$ and $b_t$, say $\hat{\mu}_t$ and $\hat{b}_t$, respectively.  The interpretation is carried out by plotting a curve of $\hat{b}_t$.  However, without elaborate construction and estimation of $b_t$, we face two issues.

The first issue is that the curve of $\hat{b}_t$ often becomes non-smooth at any time interval (i.e., every 30 minutes) in our experience.  The non-smooth daily curve of $b_t$ is unrealistic because it implies the impact of {\it daily} weather forecast on loads changes non-smoothly every 30 minutes.  This problem is caused by the fact that the regression analysis is separately conducted at each time interval.  To address this issue, we should estimate parameters under the assumption that $b_t$ is smooth.

The second issue is related to the parameter estimation procedure.  In many cases, the regression coefficients are estimated through the least squares method. However, in our experience, the estimate of regression coefficients related to $b_t$ can be negative, leading to a negative value of $\hat{b}_t$.  Since $\hat{b}_t$ is the fluctuations of electric power caused by weather, the interpretation becomes unclear.  To alleviate this problem, we need to restrict the regression coefficients associated with $b_t$ to nonnegative values.  

In this study, we develop a statistical model that elaborately captures the nonlinear structure of daily weather information to address two challenges, as mentioned earlier.  We employ the varying coefficient model \citep{Hastie:1993bx,Fan:1999ke} with basis expansion, where the regression coefficients associated with weather are assumed to be different depending on the time intervals.  The regression coefficients are expressed by a nonlinear smooth function with basis expansion, which allows us to generate a smooth function of $\hat{b}_t$.  Furthermore, the weather effect $\hat{b}_t$ is also expressed as a nonlinear smooth function.  To generate nonnegative regression coefficients, we employ the nonnegative least squares (NNLS, e.g., \citealp{doi:10.1137/1.9781611971217}) estimation.  NNLS estimates parameters under the constraint that all regression coefficients are nonnegative.  With the NNLS estimation, the value of $\hat{b}_t$ is always nonnegative; thus, the interpretation becomes clear.  

The usefulness of our proposed method is investigated through the application to three real datasets.  The results for two of the datasets show that the NNLS can appropriately capture the fluctuations of electric power caused by weather. Furthermore, our proposed method yields better forecast accuracy than the existing machine learning techniques.   In the third example, we investigate our proposed method's practical usage when COVID-19 influences the electrical loads (e.g., facility closure or recommendation of telework).  Our proposed method is directly applicable in such a situation; in addition to the daily weather forecast, we use the average number of infections in the past several days as exploratory variables.   The result shows that our proposed method can adequately capture the effect of COVID-19 and also improve forecast accuracy.

The remainder of this paper is organized as follows: Section 2 describes our proposed model based on the varying coefficient model.  In Section 3, we present the parameter estimation via nonnegative least squares. Section 4 presents the analysis of data from Tokyo Electric Power Company Holdings.  In Section 5, we investigate the impact of COVID-19 on electrical loads through the analysis of data from one selected research facility in Japan.   Concluding remarks are given in Section 6, and technical proofs are deferred to the Appendices.

\section{Proposed model}\label{sec:model}
Short- and medium-term forecasting is often used for trading electricity in the market.  Among various electricity markets, the day-ahead (or spot) and the intraday markets are popular in electricity exchanges, including the European Power Exchange (EPEX) (\url{https://www.epexspot.com/en/market-data/dayaheadauction}) and Japan Electric Power Exchange (JEPX) (\url{http://www.jepx.org/english/index.html}).   In the day-ahead market, contracts for the delivery of electricity on the following day are made.    In the intraday market, the power will be delivered several tens of minutes (e.g., 1 hour in JEPX) after the order is closed.  In both markets, transactions are typically made in 30-minute intervals; thus, the suppliers must forecast the loads in 30-minute intervals.   In this study, we consider the problem of forecasting loads that can be applied to both day-ahead and intraday markets.  

Let $y_{ij}$ be the electricity load at $j$th time interval on $i$th date ($i=1,...,n$, $j=1,...,J$).  Typically, $J=48$, because we usually forecast the loads in 30-minute intervals.  We consider the following model: 
\begin{eqnarray}
	y_{ij} = \mu_{ij} + b_{ij} + \varepsilon_{ij}, \label{eq:model}
\end{eqnarray}
where $\mu_{ij}$ is the effect of past electricity load, $b_{ij}$ is the effect of weather, such as temperature and humidity, and $\varepsilon_{ij}$ are error terms with $E[\varepsilon_{ij}]=0$.   

Typically, the error variances in the daytime are larger than those at midnight because of the uncertainty of human behavior in the daytime.  Therefore, it would be reasonable to assume that $V[\varepsilon_{ij}]=\sigma_j^2$, i.e., the error variances depend on the time interval.   One may assume the correlation of errors for different time intervals, i.e., ${\rm Cor}(\varepsilon_{ij}, \varepsilon_{ij'}) \neq 0$ for some $j \neq j'$; however, the number of parameters becomes large.  For this reason, we consider only the case where the errors are uncorrelated.  Note that the final implementation of our proposed procedure described later is independent of the assumption of the correlation structure in errors.

One can express $b_{ij}$ and $\mu_{ij}$ as linear or nonlinear functions of predictors and conduct the linear regression analysis.   With this procedure, however, we face two issues, as mentioned in the introduction; thus, we carefully construct appropriate functions of $b_{ij}$ and $\mu_{ij}$.  

\subsection{Expression of $b_{ij}$}\label{sec:b_ij}
Weather forecast information is typically observed at intervals of one day and not 30 minutes (e.g., the maximum temperature of average humidity).  For this reason, we assume that the weather forecast information does not depend on $j$.  Let a vector of weather information be $\bm{s}_{i}$.  We express $b_{ij}$ as a function of $\bm{s}_i$.  Here, we assume two structures as follows:
\begin{itemize}
    \item It is well known that the relationship between weather variables and consumption is nonlinear.   For example, the relationship between maximum temperature and consumption is approximated by a quadratic function (e.g., \citealp{Hong:2010gl}) because air conditioners are used on both hot and cold days.  For this reason, it is assumed that $b_{ij}$ is expressed as some nonlinear function of $\bm{s}_i$.
    \item  Although $\bm{s}_i$ does not depend on $j$, the effect of weather, $b_{ij}$, may depend on $j$.  For example, consumption in the daytime is affected by the maximum temperature more than that at midnight.  In this case, the regression coefficients associated with $\bm{s}_i$ change according to the time interval $j$.  However, if we assume different parameters at each time interval, the number of parameters can be large, resulting in poor forecast accuracy.  To decrease the number of parameters, we use the varying coefficient model, in which the coefficients are expressed as a smooth function of the time interval.  
\end{itemize}
Under the above considerations, we propose expressing $b_{ij}$ as follows:
\begin{eqnarray}
    b_{ij} = \sum_{m=1}^M \beta_{m}(j) g_{m}(\bm{s}_{i}), \label{eq:be}
\end{eqnarray}
where $g_{m}(\bm{s}_{i}) $  ($m=1,...,M$) are basis functions given beforehand, $\beta_m(j)$ are functions of regression coefficients, and $M$ is the number of basis functions.

 We also use the basis expansion for $\beta_{m}(j) $:
\begin{eqnarray}
	 \beta_{m}(j)  = \sum_{q=1}^Q\gamma_{qm} h_{q}(j), \label{eq:vcm}
\end{eqnarray}
where $h_{q}(j)$  ($q=1,...,Q$) are basis functions given beforehand and $\gamma_{qm}$ are the elements of the coefficient matrix $\bm{\Gamma}=(\gamma_{qm})$.  Substituting \eqref{eq:vcm} into \eqref{eq:be} results in the following:
\begin{eqnarray}
	b_{ij} = \sum_{m=1}^M \sum_{q=1}^Q\gamma_{qm} h_{q}(j) g_{m}(\bm{s}_{i}). \label{eq:bikj}
\end{eqnarray}
Because $h_{q}(j)$ and $g_{m}(\bm{s}_{i})$ are known functions, the parameters concerning $b_{ij}$ are $\gamma_{qm}$.  

Since the effect of weather is assumed to be smooth according to both $j$ and $\bm{s}_{i}$, we use basis functions $h_{q}(j)$ and $g_{m}(\bm{s}_{i})$, which produce a smooth function, such as B-spline and the radial basis function (RBF).

\subsection{Expression of $\mu_{ij}$}\label{sec:mu_ij}
Since $\mu_{ij}$ is the effect of past consumption, one can assume that $\mu_{ij}$ is expressed as a linear combination of past consumption $y_{(i-t-L_{\alpha})j}$, i.e., 
\begin{eqnarray}
    \mu_{ij} = \sum_{t=1}^T \alpha_{jt} y_{(i-t-L_{\alpha})j} + \sum_{u=1}^U \beta_{ju} y_{i(j-u-L_{\beta})},\label{eq:mu_ij naive}
    \end{eqnarray}
    where $T$ and $U$ are positive integers, which denote how far we trace back through the data and $\alpha_{jt}$ ($t=1,...,T$) and $\beta_{ju}$  ($u=1,...,U$) are positive values given beforehand.      Here, $L_{\alpha}$ and $L_{\beta}$ are nonnegative integers that describe the lags; these values change according to the closing time of transactions\footnote{For example, transactions of the day-ahead market in the JEPX close at 5:00 pm every day.  For the forecast of the 5:30--6:00 pm interval tomorrow, we cannot use the information of today's consumption at the 5:30--6:00 pm interval due to the trading hours of the market, which implies $L_{\alpha}=1$.}.  The regression coefficients $\alpha_{jt}$ correspond to the effects of past consumptions for the same time interval on previous days, while $\beta_{jt}$ are the coefficients for different time intervals on the same day.  For the day-ahead market, we assume that $\beta_{ju} \equiv 0 $.

    In practice, however, it is assumed that past consumption also depends on past weather, such as daily temperature.   For example, suppose that it was exceptionally hot yesterday and it is cooler today.  In this case, it is not desirable to directly use past consumption as the predictor; it is better to remove the effect of past temperature from past consumption.  In other words, we can use $y_{(i-t-L_{\alpha})j} - b_{(i-t-L_{\alpha})j}$ and $y_{i(j-u-L_{\beta})} - b_{i(j-u-L_{\beta})}$ instead of $y_{(i-t-L_{\alpha})j}$, and $y_{i(j-u-L_{\beta})}$, respectively.  As a result, $\mu_{ij}$ is expressed as follows: 
\begin{eqnarray}
    \mu_{ij} = \sum_{t=1}^T\alpha_{jt} (y_{(i-t-L_{\alpha})j} - b_{(i-t-L_{\alpha})j}) +  \sum_{u=1}^U  \beta_{ju} (y_{i(j-u-L_{\beta})} - b_{i(j-u-L_{\beta})}). \label{eq:mu_kj}
\end{eqnarray}
Substituting \eqref{eq:bikj} into \eqref{eq:mu_kj} results in the following:
\begin{eqnarray}
    \mu_{ij} 
    &=&  \sum_{t=1}^T \alpha_{jt} y_{(i-t-L_{\alpha})j} -   \sum_{t=1}^T  \sum_{m=1}^M \sum_{q=1}^Q\alpha_{jt}\gamma_{qm} h_{q}(j) g_{m}(\bm{s}_{i-t-L_{\alpha}}) \nonumber\\
&& +  \sum_{u=1}^U\beta_{ju} y_{i(j-u-L_{\beta})} - \sum_{u=1}^U  \sum_{m=1}^M \sum_{q=1}^Q\beta_{ju}\gamma_{qm} h_{q}(j-u-L_{\beta}) g_{m}(\bm{s}_{i}). \label{eq:mu_kj_adj}
\end{eqnarray}

The appropriate values of $\alpha_{jt}$ and $\beta_{ju}$ are chosen by several approaches.  A simple method is $\alpha_{jt}=1/T$ and $\beta_{ju}=1/U$, which implies $\mu_{ij}$ is the sample mean of the past consumption.   Another method is based on the AR(1) structure, i.e., $\alpha_{jt}=\rho^t_{\alpha}$ and $\beta_{ju}=\rho^u_{\beta}$, where $\rho_{\alpha}$ and $\rho_{\beta}$ satisfy $\sum_{t=1}^T\rho_{\alpha}^t=1$ and $\sum_{u=1}^U\rho_{\beta}^u=1$, respectively.  Note that $\sum_{t=1}^T\rho^t=\rho(1-\rho^T)/(1-\rho)$, so $\sum_{t=1}^T\rho^t=1$ is equivalent to $\rho^{T+1} -2\rho +1=0$, whose numerical solution is easily obtained.  

\subsection{Proposed model}
By combining the expressions of $b_{ij}$ in \eqref{eq:bikj}  and $\mu_{ij}$ in \eqref{eq:mu_kj_adj}, the model \eqref{eq:model} is expressed as follows:
\begin{eqnarray}
	y_{ij} &=& \frac{1}{T}\sum_{t=1}^T  y_{(i-t-L_{\alpha})j} -  \frac{1}{T}\sum_{t=1}^T  \sum_{m=1}^M \sum_{q=1}^Q\gamma_{qm} h_{q}(j) g_{m}(\bm{s}_{i-t-L_{\alpha}})\cr
&& +\frac{1}{U} \sum_{u=1}^U y_{i(j-u-L_{\beta})} - \frac{1}{U}\sum_{u=1}^U  \sum_{m=1}^M \sum_{q=1}^Q\gamma_{qm} h_{q}(j-u-L_{\beta}) g_{m}(\bm{s}_{i}) \cr
	&& + \sum_{m=1}^M \sum_{q=1}^Q\gamma_{qm} h_{q}(j) g_{m}(\bm{s}_{i})  +  \varepsilon_{ij}. \label{eq:model_adj} 
	\end{eqnarray}
	The model \eqref{eq:model_adj} is equivalent to the linear regression model
\begin{eqnarray}
	\tilde{\bm{y}} &=&\bm{X}\bm{\gamma} + \bm{\varepsilon}, \label{eq:LR}
	\end{eqnarray}
where $\bm{\gamma} = {\rm vec}(\bm{\Gamma})$ and $\bm{\varepsilon} = {\rm vec}(\bm{E})$ with $\bm{E} = (\varepsilon_{ij})$.  Here, $\bm{X}$ and $\tilde{\bm{y}}$ are considered as the design matrix and the response vector, respectively.   The definitions of $\tilde{\bm{y}}$ and $\bm{X}$ are given in \ref{app:vecmat}.

\section{Estimation}
\subsection{Nonnegative least squares}
To estimate the regression coefficient vector $\bm{\gamma}$, one can use the least squares estimation (LSE)
\begin{eqnarray*}
    \min_{\bm{\gamma}}\|\tilde{\bm{y}} -\bm{X}\bm{\gamma}\|_2^2. \label{eq:LSE}
    \end{eqnarray*}    
In our experience, however, the elements of least squares estimate $\hat{\bm{\gamma}}$ often become negative.  In such cases, the estimate of $b_{ij}$ is negative because the basis functions $h_q(j)$ and $g_m(\bm{s}_i)$ generally take positive values.  When $b_{ij} < 0$, one can interpret the weather effect is negative.   Nevertheless, the one-day curve of $\hat{b}_{ij}$ turns out to be counterintuitive; the effect of weather negatively increases as the electricity loads increase.  In other words, $b_{ij}$ is negatively large at working hours and small at midnight.  As a result,  $\mu_{ij}$ becomes extremely large in the working time.  We observe this phenomenon on the analysis of three datasets presented in this study; one of them is the well-known Global Energy Forecasting Competition 2014 (GEFCom2014) data.  Therefore, this phenomenon could occur in other datasets.    

A clear interpretation is realized when the weather effect $b_{ij}$ is nonnegative.  To achieve this, we employ the nonnegative least squares (NNLS) estimation, in which we minimize the loss function under a constraint the the regression coefficients are nonnegative:
\begin{eqnarray}
    \min_{\bm{\gamma}}\|\tilde{\bm{y}} -\bm{X}\bm{\gamma}\|_2^2 \quad {\rm subject \ to \ \ } \bm{\gamma} \geq \bm{0}. \label{eq:NNLS}
    \end{eqnarray}    
 The optimization problem \eqref{eq:NNLS} is a special case of quadratic programming with nonnegativity constraints (e.g., \citealp{Franc:2005ue}).  As a result, the NNLS problem becomes a convex optimization problem.  Several efficient algorithms to obtain the solution in  \eqref{eq:NNLS}  have been proposed in the literature (e.g., \citealp{doi:10.1137/1.9781611971217,Bro:1997el,Timotheou:2016bm}).
 
We add the ridge penalty \citep{Hoerl:1970cd} to the loss function of the NNLS estimation:
\begin{eqnarray}
    \min_{\bm{\gamma}}\|\tilde{\bm{y}} -\bm{X}\bm{\gamma}\|_2^2 + \lambda \|\bm{\gamma}\|_2^2 \quad {\rm subject \ to \ \ } \bm{\gamma} \geq \bm{0}, \label{eq:NNLSridge}
    \end{eqnarray}    
where $\lambda>0$ is a regularization parameter.  In our experience, the ridge penalization improves the forecast accuracy, and also produces $\hat{b}_{ij}$ that is easier to interpret than the unpenalized NNLS.

\subsection{Forecast}
For the day-ahead forecast,  we forecast the loads on the next day, $\hat{y}_{(i+1)j}$, for the given NNLS estimate $\hat{\bm{\gamma}}$ and weather information $\bm{s}_{i+1}$.   The forecast value $\hat{y}_{(i+1)j}$  is expressed as follows: 
\begin{equation*}
\hat{y}_{(i+1),j} = \hat{\mu}_{(i+1)j} + \hat{b}_{(i+1)j} 	
\end{equation*}
Here, $\hat{b}_{(i+1)j} = \sum_{m=1}^M \sum_{q=1}^Q\hat{\gamma}_{qm} h_{q}(j) g_{m}(\bm{s}_{i+1})$ and $\hat{\mu}_{(i+1)j} =  \sum_{t=1}^T \alpha_{jt} (y_{(i+1-t-L_{\alpha})j} - \hat{b}_{(i+1-t-L_{\alpha})j})$.   On the intraday forecast,  we may use information about the loads on that day so that   $\hat{\mu}_{(i+1)j} $ is expressed as  $$\hat{\mu}_{(i+1)j} =  \sum_{t=1}^T \alpha_{jt} (y_{(i+1-t-L_{\alpha})j} - \hat{b}_{(i+1-t-L_{\alpha})j}) + \sum_{u=1}^U \beta_{ju} (y_{(i+1)(j-u-L_{\beta})} - \hat{b}_{(i+1)(j-u-L_{\beta})}).$$

Construction of a forecast interval based on \eqref{eq:NNLS} or \eqref{eq:NNLSridge} is not easy due to the constraints of the parameter. To derive the forecast interval, we employ a two-stage procedure; first, we estimate the parameter via NNLS to extract variables that correspond to nonzero coefficients.  Then, we employ the least squares estimation based on the variables selected in the first step. With this procedure, we should derive the forecast interval after model selection.   To achieve this result, the post-selection inference \citep{Lee:2014uu,Lee:2016ila} is employed.  The post-selection inference for the NNLS estimation is detailed in \ref{app:post-selection inference}.

\section{Application to demand data from Tokyo Electric Power Company Holdings}\label{sec:toden}
The performance of our proposed method is investigated through the analysis of electricity load data collected from Tokyo Electric Power Company Holdings, available at \url{https://www.tepco.co.jp/en/forecast/html/download-e.html}.  The dataset consists of electricity loads from April 1st, 2016, to March 30th, 2020. The loads are shown in MW at 1-hour intervals (i.e., $J=24$). 

We forecast the loads from April 1st, 2019 to March 30th, 2020 (data in 2016--2018 are only used for training).  The training data consist of all load data up to the previous day of the forecast day; for example, when we forecast the loads on February 4th, 2020, the training data are loads from April 1st, 2016, to February 3rd, 2020.  We consider the problem of the day-ahead forecast, that is, $\beta_{ju} \equiv 0 $.  
\subsection{Basis functions}
We use maximum temperature in Tokyo as the weather variable $\bm{s}_i$, available at Japan Meteorological Agency (\url{https://www.jma.go.jp/jma/indexe.html}).  
\begin{figure}[!t]
\centering
\includegraphics[width=16cm]{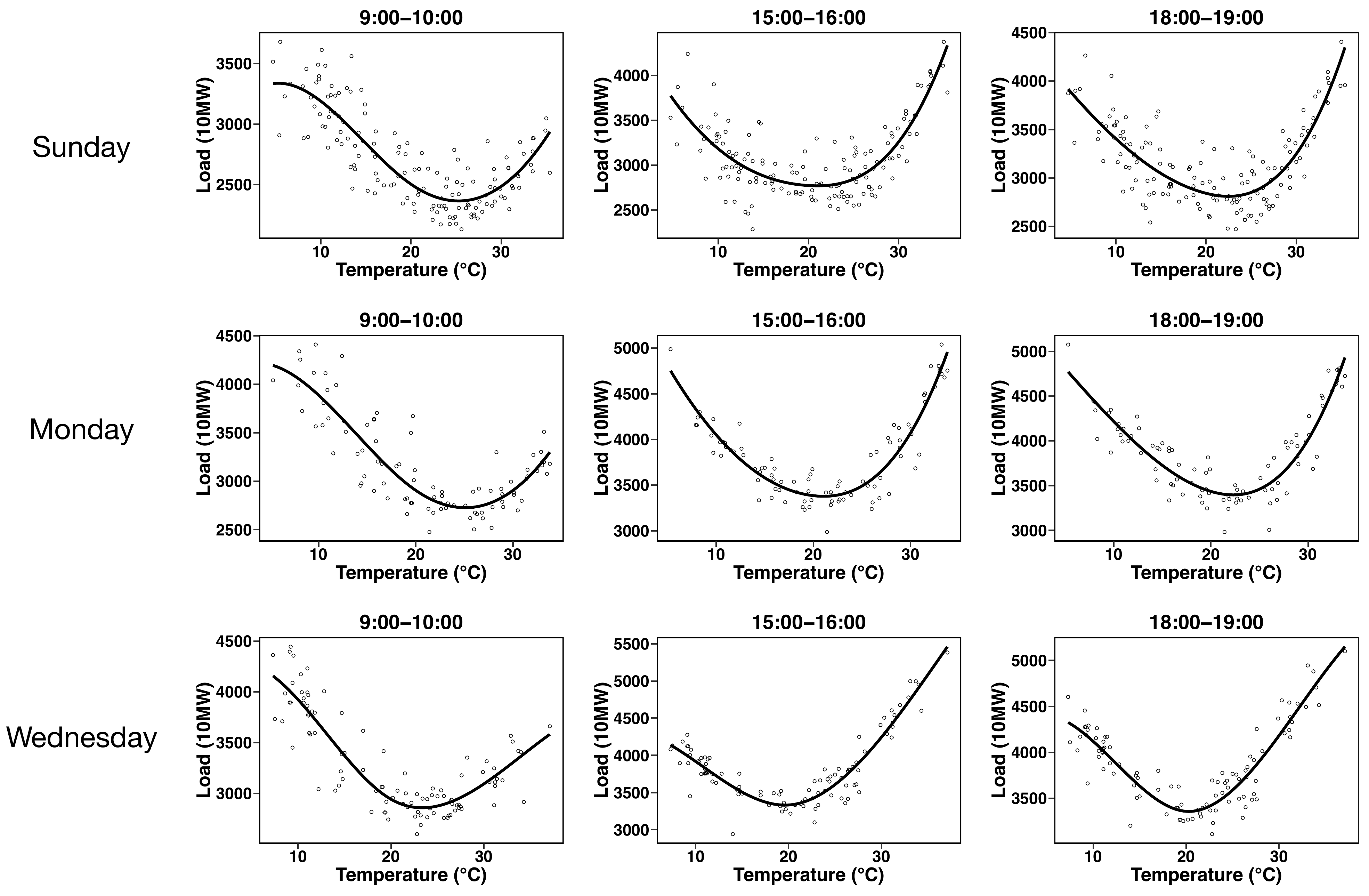}
\caption{Relationship between maximum temperature and loads for different time intervals on Sunday, Monday, and Wednesday.  The curves are depicted by fitting the ordinary least squares estimation with the cubic B-spline function.  }
\label{fig:temperatureloads_toden}
\end{figure}   
Figure \ref{fig:temperatureloads_toden} shows the relationship between the maximum temperature and loads.  We depict this relationship in different time intervals for Sunday, Monday, and Wednesday.  The curves are depicted by fitting the ordinary least squares estimation with the cubic B-spline function (e.g., \citealp{Hastieetal:2008}) with equally-spaced knots.  The number of basis functions is $M=5$. For all settings, the curve-fitting works well, which implies the usage of the B-spline function as a basis function $g(\bm{s}_i)$ would be reasonable.  

  The curve shape for 9:00--10:00 is different from that for 15:00--16:00, which suggests that the regression coefficients must be different among time intervals.  Meanwhile, the curve shapes for 15:00--16:00 and 18:00--19:00 are similar.  In this case, it is reasonable to assume that the regression coefficients for 15:00--16:00 are similar to those for 18:00--19:00.  The regression coefficients are assumed to change depending on the time interval yet to be smooth according to the time interval. To achieve a smooth function of $\beta_m(j)$, we also use the cubic B-spline as a basis function of $h_q(j)$.  As the ordinary B-spline function cannot produce a smooth curve around the boundary (i.e., 23:00--0:00 and 0:00--1:00), we employ the cyclic B-spline function, where the basis functions wrap at the first and last knot locations.  The cyclic B-spline function is implemented in the {\tt cSplineDes} function in the {\tt mgcv} package in {\tt R}.

We also observe that the curve shapes are different among each day of the week.  Therefore, we construct the statistical models by day of the week separately; seven statistical models are constructed.  To forecast the loads, we select a model that matches the day of the week. All national holidays are regarded as Sunday; therefore, the number of observations on Sunday is larger than on other weekdays.  

\subsection{Candidates of tuning parameters}

 We employ our proposed method based on two estimation procedures: ridge estimation
     \begin{eqnarray*}
    \min_{\bm{\gamma}}\|\tilde{\bm{y}} -\bm{X}\bm{\gamma}\|_2^2 + \lambda \|\bm{\gamma}\|_2^2, \label{eq:ridge}
    \end{eqnarray*}    
and NNLS estimation with the ridge penalty in \eqref{eq:NNLSridge}.  We label these estimation procedures as ``LSE" and ``NNLS," respectively.  For both LSE and NNLS, we prepare a wide variety of statistical models by changing the tuning parameters.  Here, we review the role of each tuning parameter and present their candidates as follows:
\begin{itemize}
    \item $Q$: the number of basis functions in the varying coefficient model.  As $Q$ increases, the electricity fluctuations caused by weather becomes large in time interval $j$ ($j=1,\dots,J$).  The candidates of $Q$ are $Q=5,10$.
    \item $M$: the number of basis functions for the impact of weather on loads.  As $M$ increases, the electricity fluctuations caused by weather becomes large in temperature $\bm{s}_i$.  The candidates of $M$ are $M=5,10$.
    \item $T$: the number of past loads for forecasting.  In other words, we use loads in the past $T$ days to forecast loads.  The candidates of $T$ is $T=2,4$.
    \item $\lambda$: regularization parameter for ridge regression.  As $\lambda$ increases, the regression coefficients become stable. Small $\lambda$ prevents the overfitting, but too large $\lambda$ leads to large bias.  The candidates of $\lambda$ are 20 sequences from $10^{-5}$ to 1.0 on a log-scale.  We also set $\lambda=0$ to investigate the impact of the ridge parameter on the forecast accuracy.  
\end{itemize}
In addition to the above candidates of models, we consider two types of $\alpha_{jt}$: sample mean of the past loads (i.e., $\alpha_{jt}=1/T$)  and AR(1) structure.  Details of the AR(1) structure are presented at the end of Section \ref{sec:mu_ij}.  As a result, the total number of candidates of the models is $336$ $(=2 \times 2 \times 2 \times (20+1) \times 2)$.

\subsection{Impact of tuning parameters on forecast accuracy}\label{sec:impact}
The forecast accuracy of our proposed method depends on the tuning parameters presented above.  We investigate the impact of tuning parameters on the mean average percentage error (MAPE) defined as
\begin{equation}
    {\rm MAPE}  = \frac{1}{Jn} \sum_{i=1}^n\sum_{j=1}^J \frac{|y_{ij} - \hat{y}_{ij}|}{y_{ij}}. \label{eq:MAPE}
\end{equation}

Figure \ref{fig:impacttuning} shows the  relationship  between regularization parameter $\lambda$ and MAPE.   The relationships are investigated for all combinations of $(Q, M, T)$.    
\begin{figure}[!t]
\centering
\includegraphics[width=15cm]{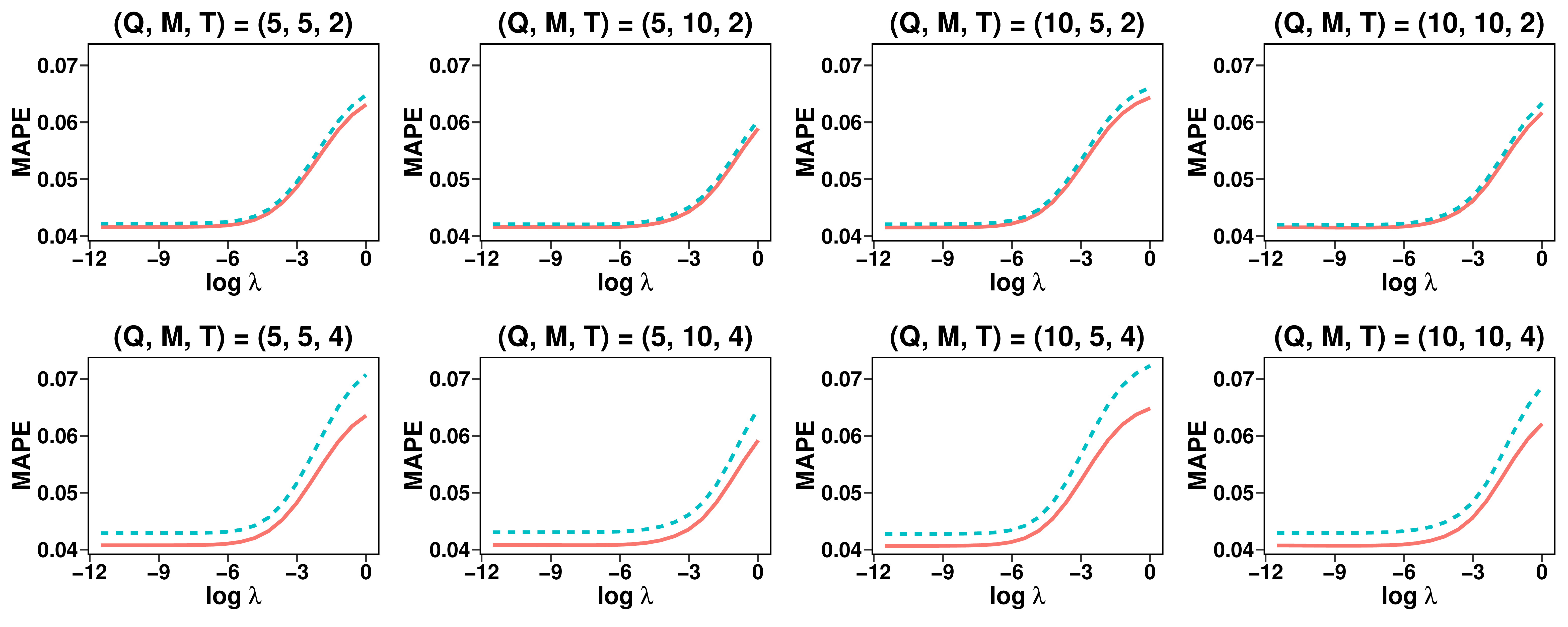}
\caption{Impact of regularization parameter $\lambda$ on the MAPE, investigated for all combinations of $(Q, M, T)$.  The dashed line indicates $\alpha_{jt} =1/T$ and the solid line corresponds to the AR(1) structure.}
\label{fig:impacttuning}
\end{figure}   
The result shows that the AR(1) structure on $\alpha_{jt}$ produces smaller MAPE than the mean structure ($\alpha_{jt}=1/T$) for all candidates of ridge parameter $\lambda$.  For AR(1) model, the performance for $T=4$ is better than that for $T=2$.  We observe that a large value of $\lambda$ may result in poor forecast accuracy; meanwhile, a small amount of $\lambda$ generally results in good accuracy. The result for $\lambda=0$ is not displayed here because the MAPE can be extremely large due to the non-convergence of parameters; the ridge penalty helps avoid such a non-convergence.    In summary, the AR(1) structure on $\alpha_{jt}$, $T=4$, and a small value of $\lambda$ will result in a small MAPE for this dataset.  

\subsection{Forecast accuracy}
We compare the performance of our proposed method with the following popular machine learning techniques: random forest (RF), support vector machine (SVM), least absolute shrinkage and selection operator (Lasso), and LightGBM (LGBM; \citealp{ke2017lightgbm}).  We use {\tt R} packages {\tt randomForest}, {\tt ksvm}, {\tt glmnet}, and {\tt lgbm} to implement these machine learning techniques.  The LightGBM is based on the gradient boosting decision tree and achieves good forecast accuracy in various fields of research including energy (e.g., \citealp{ju2019model}). For these machine learning techniques, the forecast is made by the electricity loads of past $T$ days and maximum temperature, that is, $\hat{y}_{ij}= f(s_i, y_{(i-1)j},\dots, y_{(i-T+1)j})$.  

In practice, we need to select a set of tuning parameters among candidates. The candidates of the tuning parameters for the proposed method are presented in Section \ref{sec:impact}.   For machine learning methods, the candidates are detailed in \ref{sec:tuning}.  For all methods, a set of tuning parameters is selected so that the MAPE in the past one year is minimized.  The values of tuning parameters are changed every month.

Table \ref{tab:MAPEmonthtoden} shows the monthly MAPE for our proposed method and existing methods from April 2019 to March 2020. The result shows that the proposed method performs better than existing machine learning techniques.  Although the SVM yields better performance than other existing methods in total, it yields poor performance in August (i.e., hot season). The RF and LGBM yield similar performance and these methods produce larger MAPE than our proposed method.  The lasso yields the worst performance in total, probably because it cannot capture the nonlinear relationship between temperature and loads, as in Figure \ref{fig:temperatureloads_toden}. We observe that the NNLS and LSE result in similar values of MAPE.

\begin{table}[!t]
  \caption{Monthly MAPE for our proposed methods (NNLS and LSE) and existing machine learning techniques (SVM, RF, Lasso, and LGBM) on the dataset from Tokyo Electric Power Company Holdings from April 2019 to March 2020.  The smallest MAPE is written in bold.}
  \label{tab:MAPEmonthtoden}
\centering
  \begin{tabular}{rrrrrrrrrrrrrr}
  \hline
 & Apr & May & Jun & Jul & Aug & Sep & Oct & Nov & Dec & Jan & Feb & Mar & total \\ 
  \hline
   NNLS & {\bf 4.0} & 3.7 & 3.3 & {\bf 3.9} & {\bf 4.5} & {\bf 4.8} & {\bf 3.3} & 3.2 & {\bf 4.4} & 4.2 & 4.7 & {\bf 4.9} & 4.6 \\ 
  LSE & 4.0 & {\bf 3.7} & {\bf 3.3} & 3.9 & 4.5 & 4.8 & 3.3 & {\bf 3.2} & 4.4 & {\bf 4.2} & {\bf 4.7} & 4.9 & {\bf 4.6} \\ 
  SVM & 6.1 & 4.1 & 3.4 & 6.1 & 9.6 & 7.4 & 4.9 & 4.9 & 6.8 & 5.4 & 5.6 & 6.4 & 6.4 \\ 
  RF & 6.9 & 4.3 & 3.5 & 6.9 & 8.3 & 7.3 & 5.4 & 4.8 & 6.6 & 5.6 & 5.8 & 6.4 & 6.5 \\ 
  Lasso & 8.9 & 5.8 & 4.3 & 9.0 & 11.7 & 10.5 & 8.7 & 5.5 & 7.0 & 6.7 & 7.0 & 9.2 & 8.2 \\ 
  LGBM & 6.7 & 5.1 & 4.0 & 6.8 & 8.1 & 7.5 & 5.6 & 5.1 & 6.7 & 5.7 & 6.1 & 6.6 & 6.7 \\ 
   \hline
\end{tabular}
  \end{table}

We also compute the MAPE for well-known Global Energy Forecasting Competition 2014 (GEFCom2014) data \citep{hong2016probabilistic}, and obtain a similar result as that shown in Table \ref{tab:MAPEmonthtoden}.  For detail, please refer to \ref{sec:GEFCom2014}.  

\subsection{Interpretation}
With our proposed method, the  estimated model can be interpreted by decomposing the forecast value by the effects of temperature and past loads: $\hat{y}_{ij} = \hat{b}_{ij} + \hat{\mu}_{ij}$. The values, $\hat{b}_{ij}$ and $\hat{\mu}_{ij}$, for NNLS and LSE from October 1st to 14th, 2019, are depicted in Figure \ref{fig:bmu}. 
\begin{figure}[!t]
\centering
\includegraphics[width=14cm]{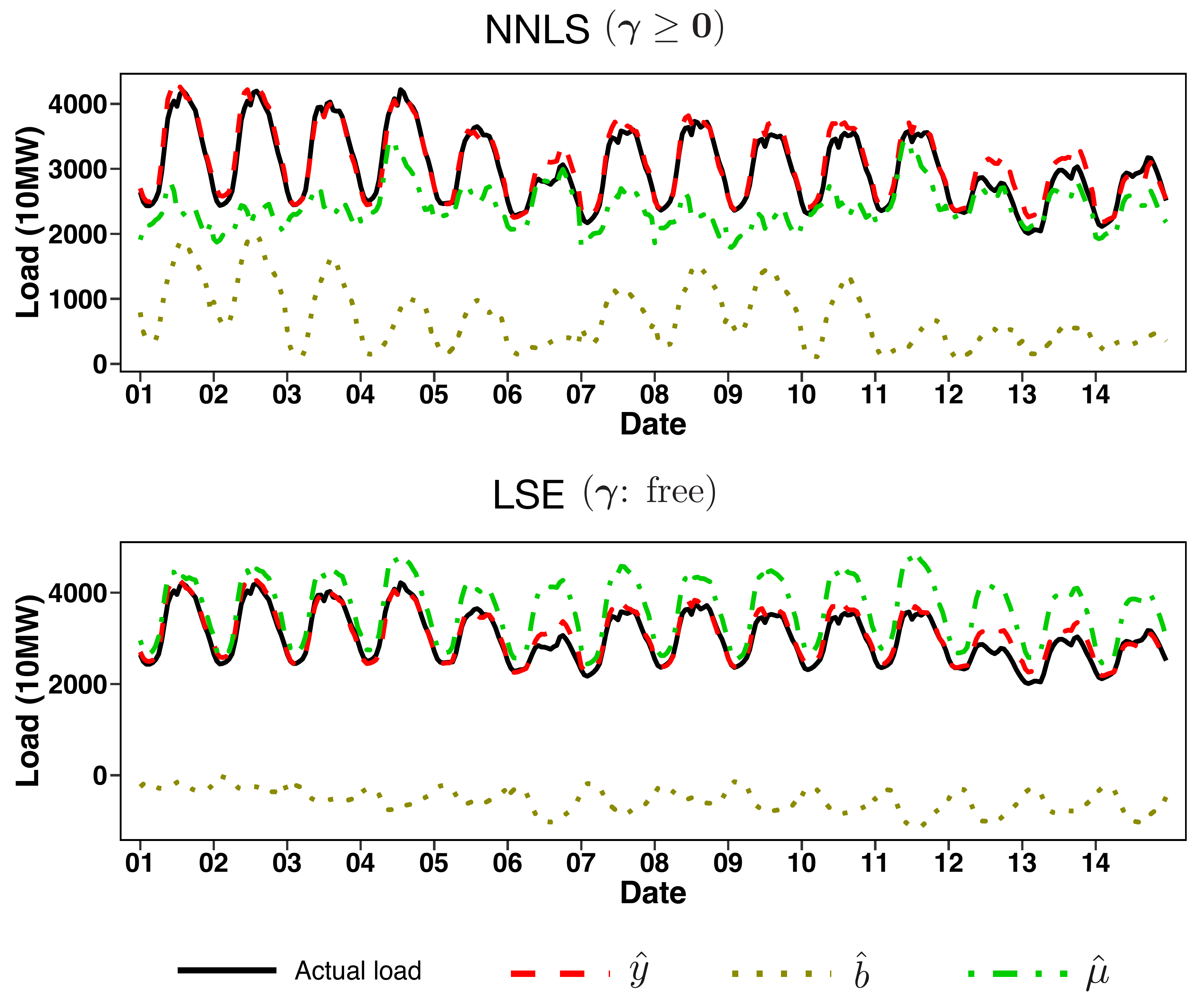}
\caption{The values, $\hat{b}_{ij}$ and $\hat{\mu}_{ij}$, for NNLS (upper panel) and LSE (lower panel) from October 1st to 14th, 2019.}
\label{fig:bmu}
\end{figure}

Although the forecast accuracy of the LSE estimation is similar to that of the NNLS estimation, as shown in Table \ref{tab:MAPEmonthtoden}, the results of the decomposition $\hat{y}_{ij} = \hat{b}_{ij} + \hat{\mu}_{ij}$ for these two methods are unalike in terms of values.  The LSE often results in negative values of $\hat{b}_{ij}$, and then $\hat{\mu}_{ij}$ becomes substantially larger than the actual load.  In particular, the value of $\hat{b}_{ij}$ at working hours is negatively larger than that at midnight; on the other hand, the electricity is used primarily during working hours.  As a result, the behavior of $\hat{b}_{ij}$ is counterintuitive; $\hat{b}_{ij}$ becomes negatively larger as $\hat{y}_{ij}$ increases. Thus, interpreting the effect of weather turns out to be difficult with LSE.  This issue occurs because there are no restrictions on the sign of $\hat{b}_{ij}$. We observe a similar behavior of $\hat{b}_{ij}$ on other datasets, including GEFCom2014.    

In contrast, the effect of weather is appropriately captured by the NNLS estimation.  Indeed, in many cases, the value of $\hat{b}_{ij}$ at working hours is positively larger than that at midnight.  The constraint on nonnegativeness of $\bm{\gamma}$ significantly improves the interpretation of the weather effect while still maintaining excellent forecast accuracy.

\begin{figure}[!t]
\centering
\includegraphics[width=15cm]{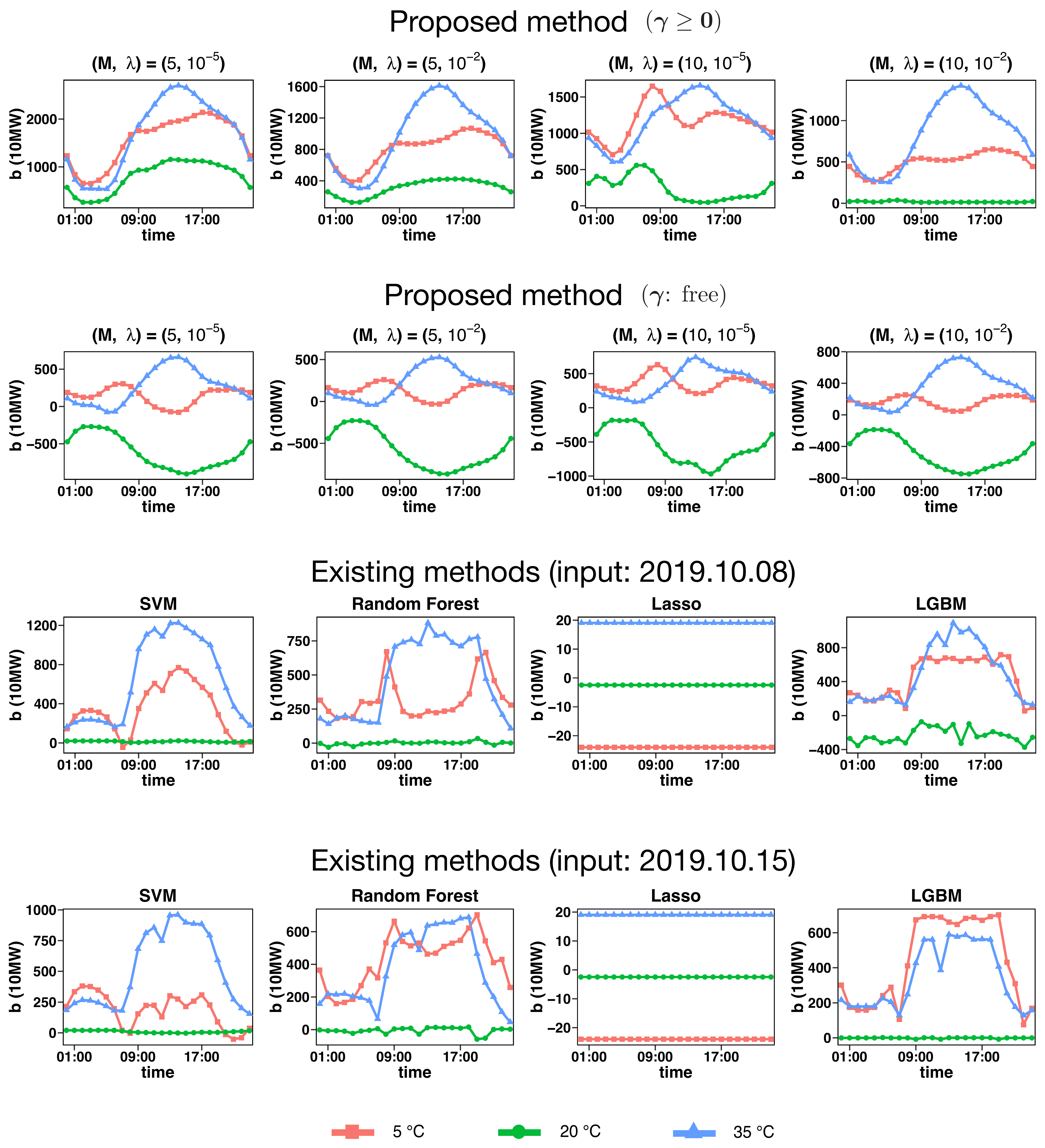}
\caption{Estimate of the weather effect $b_{ij}$ on Tuesday for the proposed method and existing methods.  For existing methods, the weather effect depends on the input.  We depict the weather effect on October 8th, 2019, and October 15th, 2019, for existing methods; both of these days are Tuesday. }
\label{fig:weathereffect}
\end{figure}   
 To further investigate the effect of temperature, we depict $\hat{b}_{ij}$ on Tuesday when maximum temperatures are 5$^\circ$C (cold day), 20$^\circ$C (cool day), and 35$^\circ$C (hot day), which is shown in Figure \ref{fig:weathereffect}.  For existing methods, it is difficult to depict $\hat{b}_{ij}$ because these methods do not assume the existence of $\hat{b}_{ij}$.   For existing methods,  instead of $\hat{b}_{ij}$,  the weather effect is calculated depending on the input: we forecast the load with specific temperature (5$^\circ$C, 20$^\circ$C, and 35$^\circ$C) and average annual temperature (21.7$^\circ$C in this case), and subtract the forecast value with average annual temperature from that with a specific temperature.  We depict the weather effect on October 8th, 2019, and October 15th, 2019, for existing methods; both of these days are Tuesday. For both our proposed method and existing methods, the parameter is estimated by using the dataset from April 1st, 2016, to October 7th, 2019.   
 
 The results of our proposed procedure show that $\hat{b}_{ij}$ is highly dependent on the temperature:  the weather effect may be substantial for cold and hot days due to air conditioner use.  For NNLS, the weather effect is always positive, which allows clear interpretation compared with LSE.  Both NNLS and LSE result in smooth curves due to the smooth basis function of the regression coefficients.  We observe that the weather effect is stable unless $M$ is large and $\lambda$ is excessively small.   
 
 For existing methods, all methods except for the Lasso are unstable and highly depend on the input.  With the Lasso, the impact of temperature is almost zero, which implies that the weather effect cannot be captured.  As a result, our proposed method is more suitable for interpreting the weather effect than the existing methods.
 
\section{Application to data from one selected research facility in Japan}
In the second real data example, we apply our proposed method to the demand data from one selected research facility in Japan. The raw data cannot be published due to confidentiality. This dataset consists of loads from January 1st, 2017 to May 30th, 2020.   At certain times, loads are either not observed or include outliers due to electricity meter failures or blackouts.  The daily data that contain such missing values and outliers are removed, resulting in 1164 days of complete data.   The loads are shown in kW at 1-hour intervals ($J=24$).  

We observe that COVID-19 greatly influences the usage of this research facility's electricity pattern due to the recommendation of telework.  In present times, it is essential to forecast the loads for this extraordinary situation. To this end, our proposed method is directly applicable to the input of daily data related to COVID-19, and we focus our attention on the forecast accuracy in March, April, and May 2020.

\subsection{Input of COVID-19 information}
 In Japan, the most frequently-used daily information about COVID-19 is the daily number of infections, available at \url{https://www3.nhk.or.jp/news/special/coronavirus/data-all/}.  However, this variable turns out to be relatively unstable.  In order to use more stable information about COVID-19, we may use the moving average of the number of infections; that is, the average number of infections in the past several days.

Figure \ref{fig:covid19_infections} shows the relationship between the number of daily infections and loads, and between the average number of infections in the past 14 days and loads.  We depict these plots on working days.  The curve fitting is done by polynomial regression with cubic function.  \begin{figure}[!t]
\centering
\includegraphics[width=16cm]{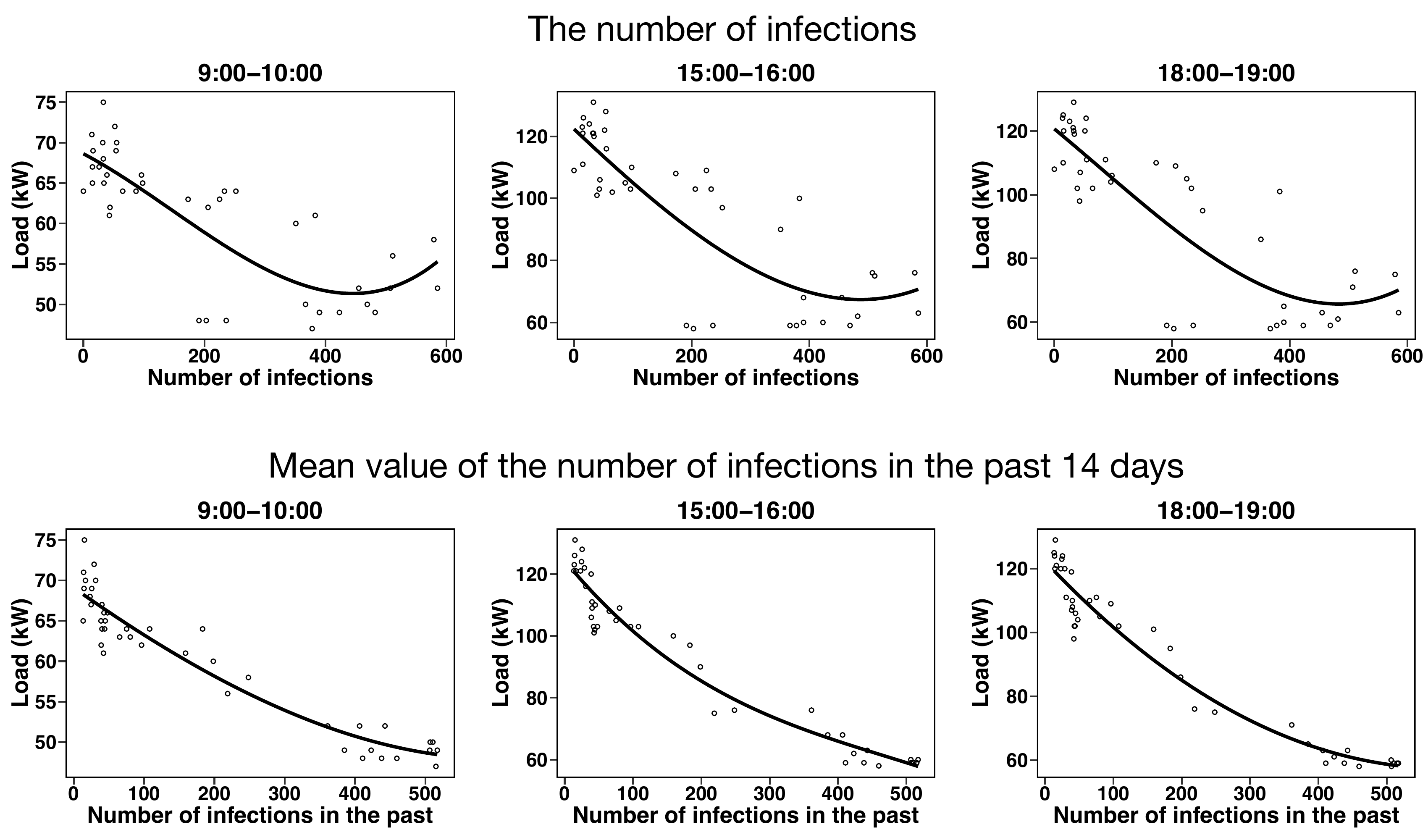}
\caption{Relationship between the number of daily infections and loads (upper panels), and between the average number of infections in the past 14 days and loads (lower panels).  We depict these plots on working days.    The curve fitting is done by polynomial regression with cubic function.}
\label{fig:covid19_infections}
\end{figure}   
The B-spline may not be suitable in this case due to the limited number of observations.  

The number of daily infections includes outliers; thus resulting in unstable curve fitting.  For example, when the number of infections is relatively large, the loads increase as the number of infections increases. This phenomenon is counterintuitive because this facility encourages working at home as much as possible to prevent infections.     On the other hand, the average number of infections in the past 14 days is negatively correlated with the loads, and the curve fitting of the cubic function works well.  For this reason, we use the average number of infections in the past 14 days as daily information about COVID-19.  As a result, the daily variable $\bm{s}_i$ becomes a two-dimensional vector that consists of the maximum temperature and the average number of infections in the past 14 days.  We use the cubic B-spline function as a basis function of maximum temperature, and the cubic function as a basis function of the average number of infections in the past 14 days.

\subsection{Forecast accuracy}
In the previous real data example, we construct the statistical models by day of the week separately. However, in this case, the number of observations affected by COVID-19 is excessively small.  Thus, only two statistical models are constructed based on working days (from Monday to Friday) and holidays (Saturday, Sunday, and national holidays).  We consider the problem of the day-ahead forecast from March 1st, 2020 to May 30th, 2020 (data in 2017--2019 are only used for training).  The training data consist of all load data up to the previous day of the forecast day.  

We investigate how the input of COVID-19 information improves the forecast accuracy.   The NNLS does not produce negative regression coefficients but we observe that the electricity usage decreases as the average number of infections in the past 14 days increases, as shown in Figure \ref{fig:covid19_infections}.  Thus, the electricity fluctuation caused by COVID-19, say $b_{ij}^{\rm covid19}$, should be negative.  As the basis function, we use the cubic function of {\it negative} number of infections in the past 14 days; that is,  $b_{ij}^{\rm covid19} = \beta_1(j) (-s_i^{\rm covid19})^3 + \beta_2(j)(-s_i^{\rm covid19})^2 + \beta_3(j)(-s_i^{\rm covid19})$, where $s_i^{\rm covid19}$ is the number of infections in the past 14 days.     

Figure \ref{fig:forecastcovid} shows the forecast values with our proposed method using NNLS and the actual loads.  We depict two forecast values: one uses the information about COVID-19 and the other does not.  The result shows that the usage of the COVID-19 data slightly improves the accuracy; for example, on the 10th and 13th.  
\begin{figure}[!t]
\centering
\includegraphics[width=15cm]{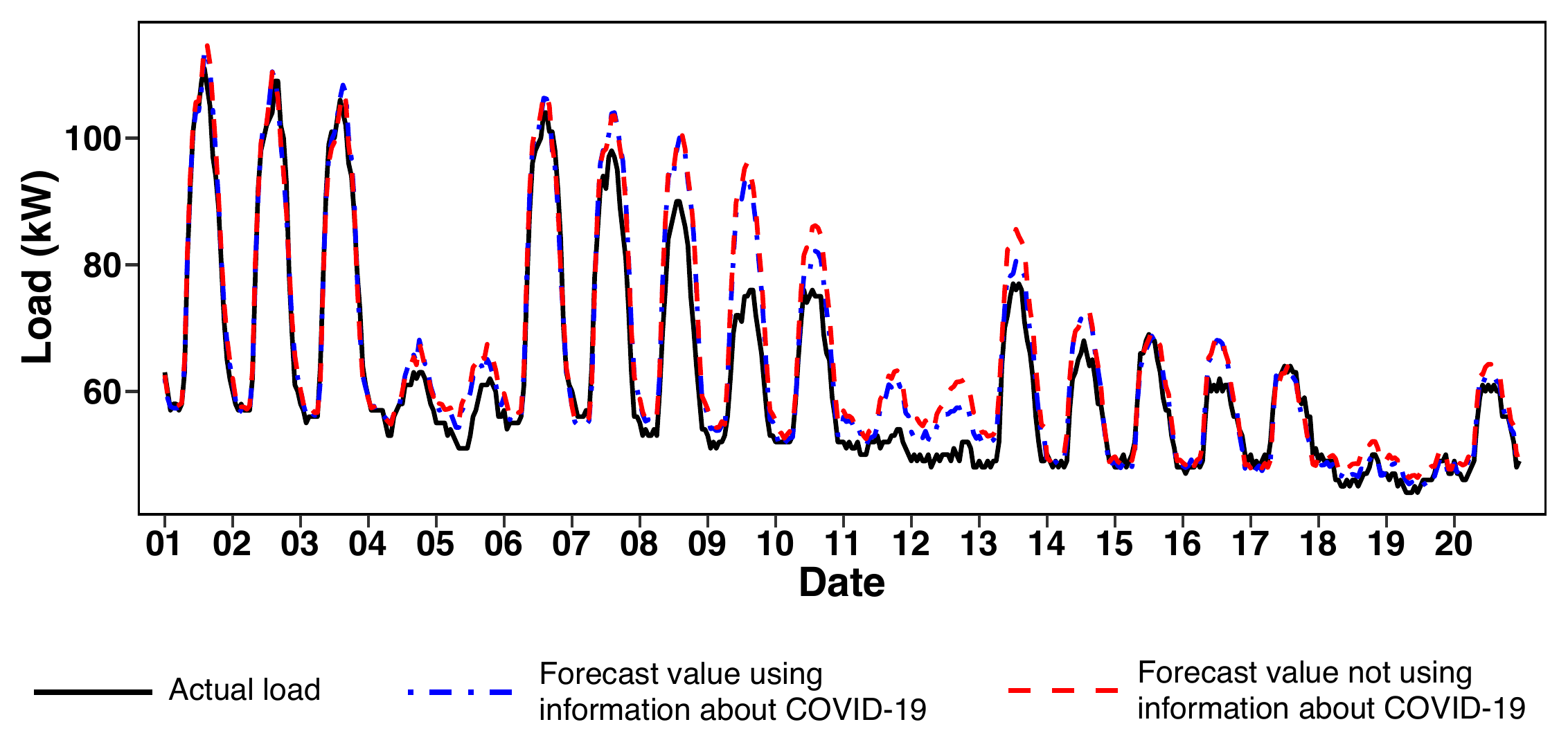}
\caption{Actual loads and forecast values from April 1st, 2020 to April 20th, 2020.  }
\label{fig:forecastcovid}
\end{figure}   

Figure \ref{fig:bhatcovid19} shows $\hat{b}_{ij}$ of both temperature and COVID-19.  The results show that after April 10th, the negative effect of COVID-19 is observed.  The government declared a state of emergency on April 7th, and following this, the usage pattern of the electricity changed. The change in electricity pattern on the 8th and 9th may not be captured due to excessively small sample sizes related to the number of infections. However, after the 9th, the effect of COVID-19 is captured.  
\begin{figure}[!t]
\centering
\includegraphics[width=15cm]{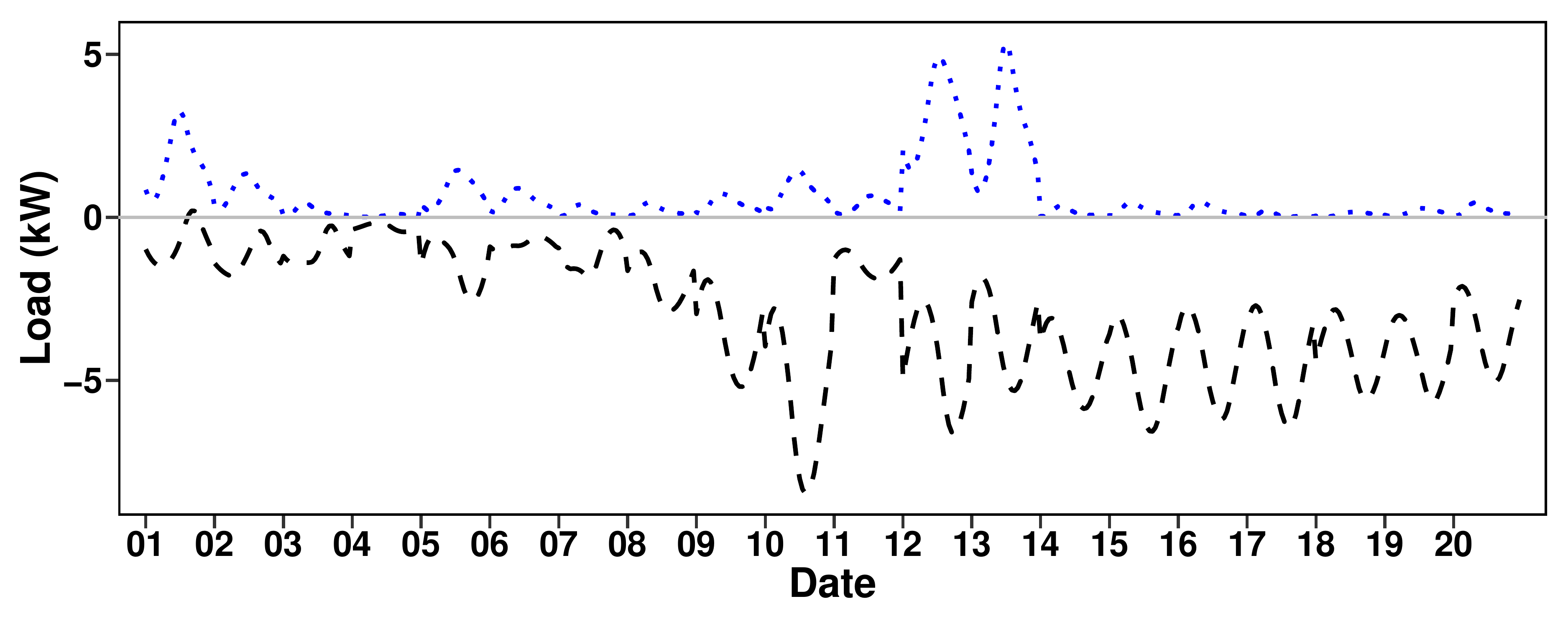}
\caption{$\hat{b}_{ij}$ for temperature (dotted line) and COVID-19 (dashed line).  }
\label{fig:bhatcovid19}
\end{figure}

Table \ref{tab:MAPEmonthcovid19} shows MAPE for our proposed method and existing methods on March, April, and May 2020.  
\begin{table}[!t]
\caption{Monthly MAPE for our proposed methods (NNLS and LSE) and existing machine learning techniques (SVM, RF, Lasso, and LGBM) on the dataset from one selected research facility in March, April, and May 2020. The smallest MAPE is written in bold.}
\label{tab:MAPEmonthcovid19}
\centering
\begin{tabular}{rrrrrrrrrrr}
  \hline
& \multicolumn{5}{c}{Including COVID-19 information} & \multicolumn{5}{c}{Not including COVID-19 information}\\
  \hline
 & NNLS & SVM & RF & Lasso & LGBM & NNLS & SVM & RF & Lasso & LGBM \\ 
  \hline
Mar & 3.5 & 5.6 & 3.9 & 4.3 & 3.9 & {\bf 3.4} & 3.8 & 3.9 & 4.1 & 4.3 \\ 
 Apr & 4.0 & 5.0 & 4.3 & {\bf 3.9} & 6.0 & 5.2 & 5.3 & 6.5 & 6.0 & 7.8 \\ 
 May & 3.1 & 3.1 & 3.5 & {\bf 2.9} & 3.4 & 3.3 & 3.2 & 3.5 & 3.4 & 3.3 \\ 
   \hline
\end{tabular}
\end{table}
We also compare the performance of two estimation procedures: one uses the information about COVID-19 and the other does not. In March, the result shows that the proposed method performs the best and Lasso yields the worst performance.  This is because the effect of temperature can be appropriately captured as shown in the previous example.  The effect of COVID-19 is not crucial in March because the performance becomes worse when the COVID-19 information is included.

In April and May, the information about COVID-19 significantly improves the performance for almost all methods.  In particular, our proposed method and Lasso both produce small MAPE values, and interestingly, Lasso performs slightly better than our proposed method.  There are two reasons to explain the superiority of Lasso. First, the relationship between the number of infections and the loads is approximated by a linear function, as shown in Figure \ref{fig:covid19_infections}.  Second, the temperature effect is not crucial because the temperature does not change much in April and May.  The nonlinear machine learning techniques, such as SVM, RF, and LGBM, yield large MAPE in April, probably due to the small number of observations related to COVID-19; these techniques result in overfitting.     As a result, only our proposed method can capture the effect of both temperature and COVID-19.

\section{Concluding remarks}
We have constructed a statistical model for forecasting future electricity loads.  To capture the nonlinear effect of weather information, we employed the varying coefficient model.  With the ordinary least squares estimation (LSE), the estimate of $b_{ij}$, say $\hat{b}_{ij}$, became negative because some of the elements of regression coefficients $\hat{\bm{\gamma}}$ were negative.  The negative weather effect led to the difficulty in the interpretation of the weather effect.  To address this issue, we employed the NNLS estimation; this estimation is performed under the constraint that all of the elements of $\hat{\bm{\gamma}}$ are nonnegative.  The practicality is illustrated through three real data examples. Two of these examples showed that our method performed better than the existing machine learning techniques.   In the third example, our proposed method adequately captured the impact of COVID-19.  Estimating the fluctuations of electric power caused by weather and COVID-19 would help make strategies for energy-saving interventions and demand response.

The proposed method is carried out under the assumption that the errors are uncorrelated.  In practice, however, the errors among near time intervals may be correlated. As a future research topic, it would be interesting to assume a correlation among time intervals and estimate a regression model that includes the correlation parameter.

\section*{Acknowledgements}
The author would like to thank Professor Hiroki Masuda and Dr. Maiya Hori for helpful comments and discussions.  This work was partially supported by the Japan Society for the Promotion of Science KAKENHI 19K11862 and the Center of Innovation Program (COI) from JST, Japan.

\appendix
\def\thesection{Appendix \Alph{section}}
\section{Matrix notation of our proposed regression model}\label{app:vecmat}
To show that our proposed model \eqref{eq:model_adj} is a regression model, we denote the following:  
\begin{flushleft}
\begin{tabular}{lp{0.6\textwidth}}
$\breve{y}_{ij\alpha} =\sum_{t=1}^T\alpha_{jt} y_{(i-t-L_{\alpha})j},$\\
$\breve{y}_{ij\beta} = \sum_{u=1}^T\beta_{jt} y_{i(j-u-L_{\beta})},$\\
$\bm{h}(j) = (h_{1}(j),\dots,h_{Q}(j))^T,$\\
$\breve{\bm{h}}_j = \sum_{u=1}^U\beta_{jt} \bm{h}(j-u-L_{\beta}),$\\
$\bm{\Gamma} = (\gamma_{qm}),$\\
$\bm{\gamma} = {\rm vec}(\bm{\Gamma}),$\\
$	\breve{g}_{m,i} = \sum_{t=1}^T\alpha_{jt} g_m(\bm{s}_{(i-t-L_{\alpha})}),$\\
$\breve{\bm{g}}_{i} = (\breve{g}_{1,i},\dots,\breve{g}_{M,i})^T,$\\
$\bm{g}_{i} = (g_1(\bm{s}_{i}), \dots, g_M(\bm{s}_{i}))^T.$\\
\end{tabular}	
\end{flushleft}
The model \eqref{eq:model_adj} is then expressed as follows: 
\begin{eqnarray*}
	y_{ij} 	&=& \breve{y}_{ij\alpha}
		 -  \bm{h}(j)^T  \bm{\Gamma} \breve{\bm{g}}_i
+ \breve{y}_{ij\beta} 
	 -  \breve{\bm{h}}_{j}^T\bm{\Gamma}\bm{g}_i 
	 +  \bm{h}(j)^T  \bm{\Gamma} \bm{g}_{i} +  \varepsilon_{ij} \label{eq:model1_element}\\
&=& \breve{y}_{ij\alpha}+ \breve{y}_{ij\beta} 
		 + \left\{ \bm{g}_{i} \otimes\bm{h}(j)-\breve{\bm{g}}_i \otimes\bm{h}(j)-\bm{g}_i \otimes\breve{\bm{h}}_{j} \right\} ^T\bm{\gamma}
+  \varepsilon_{ij}.
 \label{eq:model2_element}
	\end{eqnarray*}
Note that we use the following formula for matrices $\bm{A}$, $\bm{B}$, and $\bm{C}$:
\begin{eqnarray*}
	{\rm vec}(\bm{A}\bm{B}\bm{C}^T) &=& (\bm{C} \otimes \bm{A}) {\rm vec}(\bm{B}),\\
(\bm{A} \otimes \bm{B})^T &=& \bm{A}^T \otimes \bm{B}^T. 
\end{eqnarray*}
Furthermore, we denote the following: 
\begin{flushleft}
\begin{tabular}{lp{0.6\textwidth}}
	$\bm{q}_{i} = (\breve{y}_{i1\alpha}+ \breve{y}_{i1\beta},\dots,\breve{y}_{iJ\alpha}+ \breve{y}_{iJ\beta})^T,$\\
$\tilde{\bm{y}}_{i} = \bm{y}_{i} - \bm{q}_{i},$\\
$\tilde{\bm{y}} = (\tilde{\bm{y}}_1^T,\dots,\tilde{\bm{y}}_{n}^T)^T,$\\
$\bm{H} = ( \bm{h}(1)\dots,  \bm{h}(J)),$\\
	$	\breve{\bm{H}}= (\breve{\bm{h}}(1),\dots,\breve{\bm{h}}(J)),$\\
	$	\bm{L}_{i} = (\bm{g}_{i} \otimes\bm{H}-\breve{\bm{g}}_i \otimes\bm{H} -\bm{g}_i \otimes\breve{\bm{H}}),$\\
$\bm{L} = (\bm{L}_{1},\dots, \bm{L}_{n})^T,$\\
\end{tabular}	
\end{flushleft}
Thus, we have a linear regression model \eqref{eq:LR}.
\section{Post-selection inference for the NNLS estimation}\label{app:post-selection inference}
\subsection{Selection event of NNLS}
In this section, $\tilde{\bm{y}}$ and $\bm{\gamma}$ are referred to as  $\bm{y}$ and $\bm{\beta}$, respectively, which leads to a standard notation of the linear regression model:
\begin{equation*}
	\bm{y} = \bm{X}\bm{\beta}  + \bm{\varepsilon}.
\end{equation*}  
The post-selection inference for the NNLS estimation has already been proposed by \cite{Lee:2014uu}, but these authors lack a parameter constraint.  We have added a constraint on the parameter of the selection event.  Let $\hat{S}$ be indices that correspond to variables selected on the basis of the NNLS estimation, i.e., $\hat{S}=\{j \mid \hat{\beta}_j \neq 0 \}$.  Let $-\hat{S}$ be indices of variables not selected in $\hat{S}$.  The KKT condition in the NNLS estimation \citep{Franc:2005ue,Chen:2011ei} is
\begin{eqnarray}
	\frac{\partial L(\bm{\beta},\bm{\mu})}{\partial \hat{\bm{\beta}}}  = X^TX\hat{\bm{\beta}} -X^T\bm{y} -\bm{\mu} &=& \bm{0},\label{eq:KKT1-1}\\
	\hat{\bm{\beta}} &\geq &\bm{0},\label{eq:KKT1-2}\\
	\bm{\mu} &\geq &\bm{0},\label{eq:KKT1-3}\\
	\hat{\bm{\beta}}^T\bm{\mu} &=& 0,\label{eq:KKT1-4}
\end{eqnarray}
where $\bm{\mu}$ is the Lagrange multiplier.  Substituting $\hat{\bm{\beta}} = (\hat{\bm{\beta}}_{\hat{S}}^T,\hat{\bm{\beta}}_{-\hat{S}}^T)^T$ into \eqref{eq:KKT1-1} -- \eqref{eq:KKT1-4} results in 
\begin{eqnarray}
	-X_{\hat{S}}^T(\bm{y} - X_{\hat{S}}\hat{\bm{\beta}}_{\hat{S}}) &\geq& \bm{0}, \label{eq:KKT2-1}\\
	-X_{-\hat{S}}^T(\bm{y} - X_{\hat{S}}\hat{\bm{\beta}}_{\hat{S}}) &\geq& \bm{0},\label{eq:KKT2-2}\\
	\hat{\bm{\beta}}_{\hat{S}} &>&\bm{0} ,\label{eq:KKT2-3}\\
	\hat{\bm{\beta}}_{-\hat{S}} &=& \bm{0},\label{eq:KKT2-4}\nonumber\\
	\hat{\bm{\beta}}_{\hat{S}}^TX_{\hat{S}}^T(\bm{y} - X_{\hat{S}}\hat{\bm{\beta}}_{\hat{S}})&=&0.\label{eq:KKT2-5}
\end{eqnarray}
By combining \eqref{eq:KKT2-1}, \eqref{eq:KKT2-3}, and \eqref{eq:KKT2-5}, we obtain
\begin{eqnarray}
	X_{\hat{S}}^T(\bm{y} - X_{\hat{S}}\hat{\bm{\beta}}_{\hat{S}}) &=& \bm{0}.\label{eq:KKT3-1}
\end{eqnarray}
Then, we obtain
\begin{eqnarray}
	\hat{\bm{\beta}}_{\hat{S}} = (X_{\hat{S}}^TX_{\hat{S}})^{-1}X_{\hat{S}}^T\bm{y}.\label{eq:KKT3-2}
\end{eqnarray}
Eq. \eqref{eq:KKT3-2} implies the NNLS estimate for the active set coincides with the LSE using the active set.  Substituting \eqref{eq:KKT3-2}  into \eqref{eq:KKT2-2} and \eqref{eq:KKT2-3} results in 
\begin{eqnarray}
	-X_{-\hat{S}}^T(\bm{I} - X_{\hat{S}}(X_{\hat{S}}^TX_{\hat{S}})^{-1}X_{\hat{S}}^T)\bm{y} &>& \bm{0},\label{eq:KKT2}\\
	(X_{\hat{S}}^TX_{\hat{S}})^{-1}X_{\hat{S}}^T\bm{y}&\geq&\bm{0}. \label{eq:KKT3}
\end{eqnarray}
 The selection event constructed by \eqref{eq:KKT2} and \eqref{eq:KKT3} is expressed as
\begin{eqnarray*}
\hat{E}(\bm{y}) &=& \{ \bm{y} \mid A(\hat{S}) \geq0\},
\end{eqnarray*}
where $A(\hat{S})$ is given by 
\begin{eqnarray*}
	A(\hat{S}) = 
	\begin{pmatrix}
X_{-\hat{S}}^T(\bm{I} - X_{\hat{S}}X_{\hat{S}}^{\dagger} )\\
-X_{\hat{S}}^{\dagger} 
	\end{pmatrix}
	.
	\label{eq:As}
\end{eqnarray*}
This selection event for the NNLS estimation has already been studied by \citet{Lee:2014uu} but \citet{Lee:2014uu} did not include the first inequality \eqref{eq:KKT2}.   
\subsection{Distribution of the forecast value after model selection}
Suppose that $\bm{y} \sim N(\bm{\mu},\sigma^2I)$, and consider the problem of deriving the following distribution:
$$\bm{\eta}^T\bm{y} \mid \{ A\bm{y} \leq \bm{b} \}.$$
Here, $\bm{\eta}$ is an $n$-dimensional vector given beforehand.  For example, if $\bm{\eta}=X_{\hat{S}}^{\dagger T}\bm{e}_j$, then $\bm{\eta}^T\bm{y} = \hat{\beta}_j$.   If
\begin{eqnarray*}
	\bm{c} &=& \bm{\eta}(\bm{\eta}^T\bm{\eta})^{-1},\\
	\bm{z} &=& (I_n - \bm{c}\bm{\eta}^T)\bm{y},
\end{eqnarray*}
then
\begin{eqnarray*}
	\bm{\eta}^T\bm{y} \mid \{ A\bm{y} \leq \bm{b}, \bm{z} = \bm{z}_0 \} \sim {\rm TN}(\bm{\eta}^T\bm{\mu},\sigma^2\|\bm{\eta}\|^2,\mathcal{V}^-(\bm{z}_0),\mathcal{V}^+(\bm{z}_0)),
\end{eqnarray*}
where 
\begin{eqnarray*}
	\mathcal{V}^-(\bm{z}) &=& \max_{j:(A\bm{c})_j < 0}\frac{b_j - (A\bm{z})_j}{(A\bm{c})_j},\\
	\mathcal{V}^+(\bm{z}) &=& \min_{j:(A\bm{c})_j > 0}\frac{b_j - (A\bm{z})_j}{(A\bm{c})_j}.
\end{eqnarray*}
However, we only observe the distribution of $\bm{\eta}^T\bm{y}$ for a given $\bm{z}$.  We consider the marginalization with respect to $\bm{z}$.  To explain this result, we consider the distribution of a truncated normal distribution 
\begin{eqnarray*}
	F_{\mu,\sigma^2}^{[a,b]}(x)  = \frac{\Phi((x-\mu)/\sigma) - \Phi((a-\mu)/\sigma) }{\Phi((b-\mu)/\sigma) - \Phi((a-\mu)/\sigma)}.
\end{eqnarray*}
Then, we have
\begin{eqnarray*}
	F_{\bm{\eta}^T\bm{\mu},\sigma^2\|\bm{\eta}\|^2}^{[\mathcal{V}^{-(\bm{z})},\mathcal{V}^{+(\bm{z})}]} (\bm{\eta}^T\bm{y}) \mid \{A\bm{y} \leq \bm{b}\} \sim U(0,1),
\end{eqnarray*}
which implies
\begin{eqnarray*}
	P \left( \frac{\alpha}{2} \leq F_{\bm{\eta}^T\bm{\mu},\sigma^2\|\bm{\eta}\|^2}^{[\mathcal{V}^{-(\bm{z})},\mathcal{V}^{+(\bm{z})}]} (\bm{\eta}^T\bm{y}) \leq  1 - \frac{\alpha}{2} \mid A\bm{y} \leq \bm{b} \right) = 1 - \alpha.
\end{eqnarray*}
To construct the confidence interval for a given new input $\bm{x}$, we let $\bm{\eta}=X_{\hat{S}}^{\dagger T}\bm{x}$, and we find $L$ and $U$, which satisfies the following equation:
\begin{eqnarray*}
F_{L,\sigma^2\|\bm{\eta}\|^2}^{[\mathcal{V}^{-(\bm{z})},\mathcal{V}^{+(\bm{z})}]} (\bm{\eta}^T\bm{y})  = \frac{\alpha}{2}, \quad F_{U,\sigma^2\|\bm{\eta}\|^2}^{[\mathcal{V}^{-(\bm{z})},\mathcal{V}^{+(\bm{z})}]} (\bm{\eta}^T\bm{y})  = 1-\frac{\alpha}{2}.
\end{eqnarray*}
Letting $\bm{y}^* = (\bm{y}^T, \varepsilon)^T$ with $\varepsilon \sim N (0,\sigma^2)$ and $\eta = (\bm{x}^T,1)^T$, we construct the prediction interval.    The algorithm of the post-selection inference for the NNLS estimation procedure is shown in Algorithm \ref{algorithm:NNLS}.   

\begin{algorithm}[!t]
\caption{Selective inference for NNLS}\label{algorithm:NNLS}
\begin{algorithmic}[1]
\STATE Conduct NNLS, and obtain a set of indices for the nonzero coefficients $\hat{S}$.
\STATE Calculate $X_{\hat{S}}^{\dagger} = (X_{\hat{S}}^TX_{\hat{S}})^{-1}X_{\hat{S}}^T$.
\STATE Define the matrix of selection event $	A(\hat{S}) = 
	\begin{pmatrix}
X_{-\hat{S}}^T(\bm{I} - X_{\hat{S}}X_{\hat{S}}^{\dagger} )\\
-X_{\hat{S}}^{\dagger} 
	\end{pmatrix}
$. 
\FOR{$j = 1,\dots,|\hat{S}|$}
\STATE Let $\bm{\eta}=X_{\hat{S}}^{\dagger T}\bm{e}_j$.
\STATE Calculate $	\bm{c} = \Sigma\bm{\eta}(\bm{\eta}^T\Sigma\bm{\eta})^{-1},$ $\bm{z} = (I_n - \bm{c}\bm{\eta}^T)\bm{y}$.\vspace{3mm}
\STATE Calculate $\displaystyle \mathcal{V}^-(\bm{z}) =\max_{j:(A\bm{c})_j < 0}\frac{- (A\bm{z})_j}{(A\bm{c})_j},$ $\displaystyle \mathcal{V}^+(\bm{z}) =\min_{j:(A\bm{c})_j > 0}\frac{- (A\bm{z})_j}{(A\bm{c})_j}$.\vspace{3mm}
\STATE Find $L$ and $U$, which satisfies $\displaystyle 
F_{L,\bm{\eta}^T\Sigma\bm{\eta}}^{[\mathcal{V}^{-(\bm{z})},\mathcal{V}^{+(\bm{z})}]} (\bm{\eta}^T\bm{y})  = \frac{\alpha}{2}$ and $\displaystyle F_{U,\bm{\eta}^T\Sigma\bm{\eta}}^{[\mathcal{V}^{-(\bm{z})},\mathcal{V}^{+(\bm{z})}]} (\bm{\eta}^T\bm{y})  = 1-\frac{\alpha}{2}.
$
\ENDFOR
\end{algorithmic}
\end{algorithm}

\section{Tuning parameter for existing methods}\label{sec:tuning}
For all existing methods, we use $T=2$ and $T=4$.   Tuning parameters in existing machine learning techniques are summarized as follows
\begin{itemize}
	\item In SVM, we use the Gaussian Kernel as a Kernel function. In this case, we have three tuning parameters: $\sigma$, $C$, and   $\epsilon$. $\sigma$ is used in the Gaussian Kernel given by $K(\bm{x},\bm{x}') = \exp(-\sigma \| \bm{x} - \bm{x}'\|^2)$, and $C$ corresponds to the regularization parameter in the SVM problem.  $\epsilon$ is used in the regression version of SVM; data in $\epsilon$-tube around the prediction value is not penalized. The candidates for these parameters are $\sigma = 0.001,0.01,0.1$, $C= 1,10,100$ and $\epsilon=0.001,0.01,0.1$.  
	\item The tuning parameters of random forest include the number of trees, $n_{trees}$, and the number of variables sampled at each split, $n_{var}$. The candidates of these parameters are $n_{trees} = 50,100,500$ and $n_{var} = 1,2,3$. 
	\item In the lasso, the tuning parameter is the regularization parameter, $\lambda$.  The candidates of regularization parameter $\lambda$ are set as follows:  maximum and minimum values of $\lambda$ are defined by $\lambda_{\max}=200$ and $\lambda_{\min}=0.01$, respectively, and 100 grids are constructed on a log scale.  
	\item  For LGBM, we change two tuning parameters: the number of boosting iterations, $n_{iterations}$, and the learning rate, LR.  We set candidates of these parameters as  $n_{iterations} = 50, 100, 500,$ and $LR = 0.01,0.05,0.1,0.2,0.5,1$.   Other parameters are set to be default of the LightGBM package in \url{https://github.com/microsoft/LightGBM/tree/master/R-package}.  
\end{itemize}

\section{Application to GEFCom2014 data}
\label{sec:GEFCom2014}
We apply our method to the Global Energy Forecasting Competition 2014 (GEFCom2014) data \citep{hong2016probabilistic}, available at \url{http://blog.drhongtao.com/2017/03/gefcom2014-load-forecasting-data.html}.  The dataset consists of electricity loads from January 1st, 2006, to December 31th, 2014. The loads are shown in MW at 1-hour intervals (i.e., $J=24$). The temperature in 1-hour intervals is available, but we only use the maximum temperature as inputs.  The forecasting strategies, including the tuning parameter selection, are the same as described in Section \ref{sec:toden}.

We forecast the loads from January 1st, 2012 to December 31th, 2014.  Table \ref{tab:MAPEmonthGEFCom2014} shows the monthly MAPE for our proposed methods and machine learning techniques.  Each value represents the average value of the MAPEs in three years.    The result is similar to that of Tokyo Electric Power Company Holdings in Table \ref{tab:MAPEmonthtoden}; our proposed method performs better than existing methods, and LSE and NNLS result in similar MAPE.  

\begin{table}[!t]
\caption{Monthly MAPE for our proposed methods (NNLS and LSE) and existing machine learning techniques (SVM, RF, Lasso, and LGBM) on the dataset from Global Energy Forecasting Competition 2014.  Each value represents the average value of the MAPEs in 2012--2014. The smallest MAPE is written in bold.}
\centering
\label{tab:MAPEmonthGEFCom2014}
\begin{tabular}{rrrrrrrrrrrrrr}
  \hline
 & Jan & Feb & Mar & Apr & May & Jun & Jul & Aug & Sep & Oct & Nov & Dec & total \\ 
  \hline
NNLS& 3.4 & 2.9 & {\bf 3.0} & {\bf 2.6} & {\bf 3.0} & {\bf 4.1} & 4.9 & {\bf  4.0} & 4.9 & 2.1 & 3.8 & {\bf 3.9} & 3.5 \\ 
LSE & {\bf 3.4} & {\bf  2.9} & 3.0 & 2.6 & 3.0 & 4.1 & {\bf  4.9} & 4.0 & {\bf 4.9} & {\bf 2.1} & {\bf 3.7} & 3.9 & {\bf 3.5} \\ 
  SVM & 4.5 & 4.1 & 4.1 & 3.6 & 3.5 & 4.6 & 6.4 & 5.3 & 5.7 & 2.6 & 5.2 & 5.1 & 4.6 \\ 
  RF & 4.5 & 4.1 & 4.2 & 3.7 & 3.5 & 5.0 & 6.2 & 5.3 & 6.3 & 2.5 & 4.9 & 5.0 & 4.6 \\ 
  Lasso & 6.1 & 4.3 & 4.7 & 5.1 & 4.1 & 6.4 & 8.7 & 6.5 & 9.0 & 2.8 & 5.4 & 5.9 & 5.8 \\ 
  LGBM & 4.4 & 4.2 & 4.3 & 3.7 & 3.6 & 5.0 & 6.0 & 5.4 & 6.0 & 2.7 & 5.1 & 5.0 & 4.6 \\ 
   \hline
\end{tabular}
\end{table}

\bibliographystyle{abbrvnat}
\bibliography{paper-ref}

\end{document}